\documentclass[preprint]{elsarticle}
\usepackage{graphicx}
\usepackage{dcolumn}
\usepackage{bm}
\begin{document}

\def\vri{\vec{r}_{i}}
\def\vrj{\vec{r}_{j}}
\def\rij{r_{ij}}
\def\vrij{\vec{r}_{ij}}
\def\drij{\hat{r}_{ij}}
\def\vdr{\delta\vec{r}}
\def\dr{\delta{r}}
\def\s{\hat{s}}
\def\vrij{\vec{r}_{ij}}
\def\drij{\hat{r}_{ij}}
\def\vdr{\delta\vec{r}}
\def\dr{\delta{r}}
\def\s{\hat{s}}

\title{Jamming in two-dimensional packings }



\author[rvt]{Sam Meyer}
\author[focal]{Chaoming Song}
\author[focal]{Yuliang Jin}
\author[focal]{Kun Wang}
\author[focal]{Hern\'an A. Makse\corref{cor1}}
\ead{hmakse@lev.ccny.cuny.edu}
\cortext[cor1]{Corresponding author}
\address[rvt]{\'{E}cole Normale Sup\'erieure de Lyon,
  Universit\'{e} Claude Bernard, Lyon I, France 695014}
\address[focal]{Levich Institute and Physics Department, City College of New
  York, New York, NY, USA 10031}

\begin{abstract}
    We investigate the existence of random close and
    random loose packing limits in two-dimensional packings of
    monodisperse hard disks. A statistical mechanics approach--- based
    on several approximations to predict the probability distribution
    of volumes--- suggests the existence of the limiting densities of
    the jammed packings according to their coordination number and
    compactivity.  This result has implications for the understanding
    of disordered states in the disk packing problem as well as the
    existence of a putative glass transition in two dimensional
    systems.
\end{abstract}


\maketitle

\section{Introduction}


The concept of jamming is a common feature of out of equilibrium
systems experiencing a dynamical arrest ranging from emulsions,
colloids, glasses and spin glasses, as well as granular materials
\cite{coniglio}.  For granular matter, it is argued that a statistical
mechanical description can be used, with volume replacing energy as
the conservative quantity \cite{sirsam}. In this framework, a
mesoscopic model has been presented \cite{swm}, allowing the
development of a thermodynamics for jamming in any dimension. Here we
develop this theoretical approach to investigate the existence of
disordered packings in two-dimensional systems composed of equal-sized
hard disks \cite{berryman}. The existence of amorphous packings in 2d
is a problem of debate in the literature: two dimensional systems are
found to crystallize very easily since disordered packings of disks
are particularly unstable \cite{wyart}.

In two dimensional Euclidean space,
the hexagonal packing arrangement of circles (honeycomb circle
packing) has the highest density of all possible plane packings
(ordered or disordered) with a volume fraction $\phi_{\rm hex} =
\frac{\pi}{\sqrt{12}} \simeq 0.9069$ and each disk surrounded by 6
disks.
Regarding amorphous packings, experiments find a maximum density of
random close packing (RCP) of monodisperse spheres at $\phi_{\rm
  rcp}\approx 0.82$ \cite{berryman} while the lower limit (random
loose packing, RLP) has been little investigated and its existence has
not been treated so far.  The tendency of 2d packings to easily
crystallize has led to consider bidisperse systems which pack at a
higher RCP volume fractions of $\phi_{\rm rcp}\approx 0.84$
\cite{ohern,kertesz}.

In parallel to studies in the field of jamming--- which consider the
packing problem as a jamming transition approached from the solid
phase \cite{makse,ohern,kertesz}--- other studies attempt to
characterize jamming approaching the transition from the liquid phase
\cite{parisi,stillinger-torquato}.
Here, amorphous jammed packings are seen as infinite pressure glassy
states \cite{parisi}.  Therefore, the existence of disordered jammed
structures (of frictionless particles) is related to the existence of
a glass transition in 2d \cite{parisi}, a problem that has been
debated recently \cite{glass1}.


Here we treat the disordered disk packing problem with the statistical
mechanics of granular jammed matter \cite{sirsam}. The formal analogy
with classical statistical mechanics is the following: the
microcanonical ensemble, defined by all microstates with fixed energy,
is replaced by the ensemble of all jammed microstates with fixed
volume. Hence, the appropriate function for the description of the
system is no longer the Hamiltonian, but the \emph{volume function},
${\cal W}_i$, giving the volume available to each particle unit such that the
total system volume is $V= \sum_i {\cal W}_i$ \cite{sirsam}.

The aim of the present work is to develop the model presented in
\cite{swm} for the calculation of the volume function in the case of
2d packings.  The validity of the hypothesis employed in \cite{swm}
are discussed, and they are modified according to the properties of 2d
packings.
We use our results to study the nature of the RLP and RCP limit in 2d
through an elementary construction of a statistical mechanics that
allows the study of the existence of a maximum and minimum attainable
density of disordered circle packings. We find that amorphous packings
can pack between the density limits of $\sim$77.5\% and $\sim$80.6\%
defining the RLP and RCP respectively, according to system
coordination number and friction, opening such predictions to
experimental and computational investigation. While these values
should be considered as bounds to the real values due to the
approximations used in the theory, they serve to suggest the existence
of both limits in two dimensional packings of monodisperse disks.

It should be noted that this theoretical model is developed
for disordered packings, and RLP and RCP represent two well
defined bounds in the model under isostatic conjecture. However,
the nature of RCP in 2d is still not clear. A recent study
\cite{jin} has shown that partially crystallized jammed packings
exist in 3d and RCP could be interpreted as the ``freezing point"
in a first-order phase transition between ordered and disordered
packing phases. It is possible that a similar first-order phase
transition exists in 2d as well \cite{shattuck, radin2d}, or that
there is a continuous variation at RCP, since crystallization of
two dimensional systems can be easily achieved. Beyond the pure
amorphous packings, crystallized states should be taken into
account and future work is still required to complete the picture
in 2d.

\section{Volume function}

\begin{figure}[h]
\centerline{(a) \includegraphics[width=0.55\linewidth]{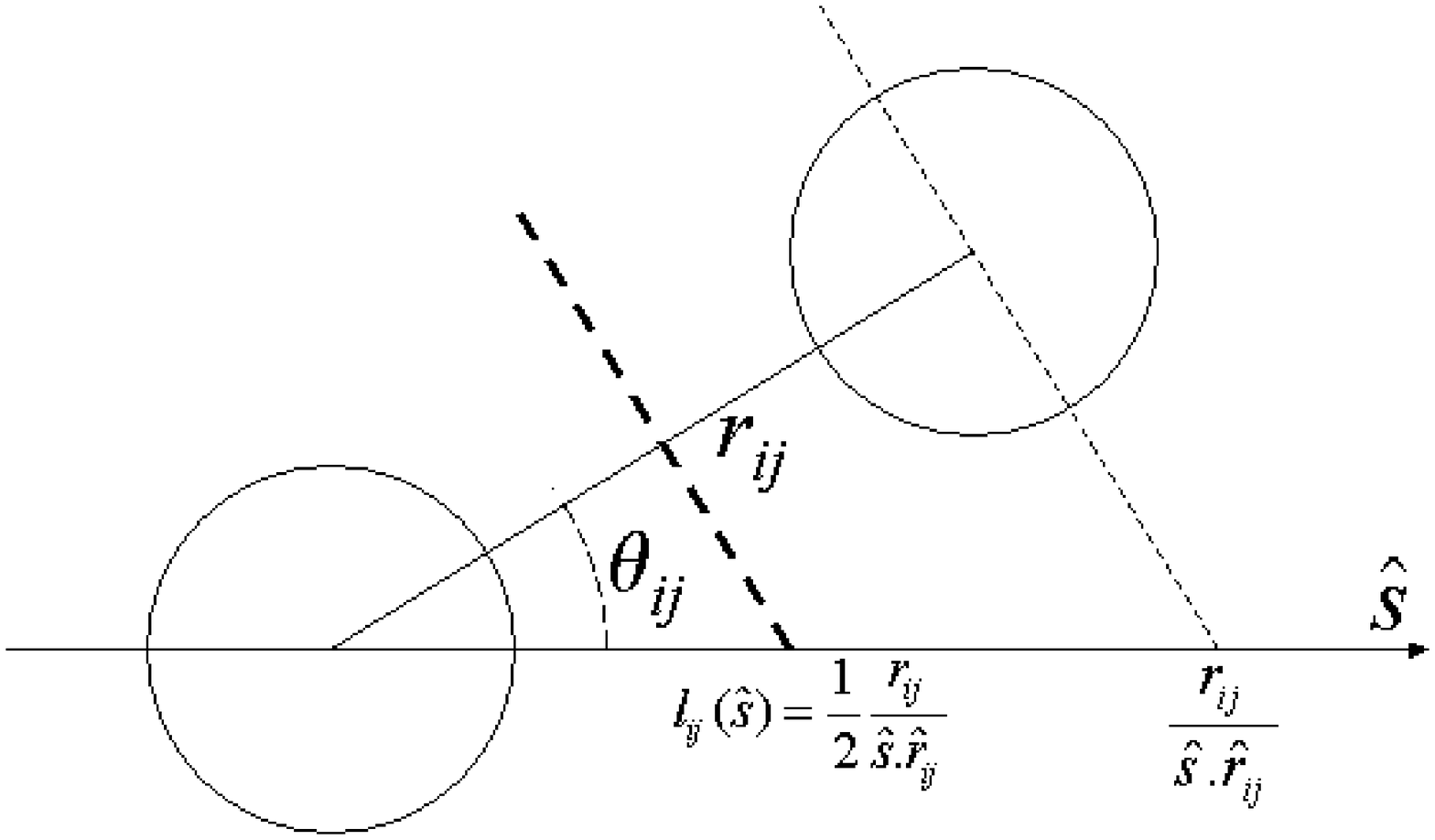}
}
\centerline{(b) \includegraphics[width=0.55\linewidth]{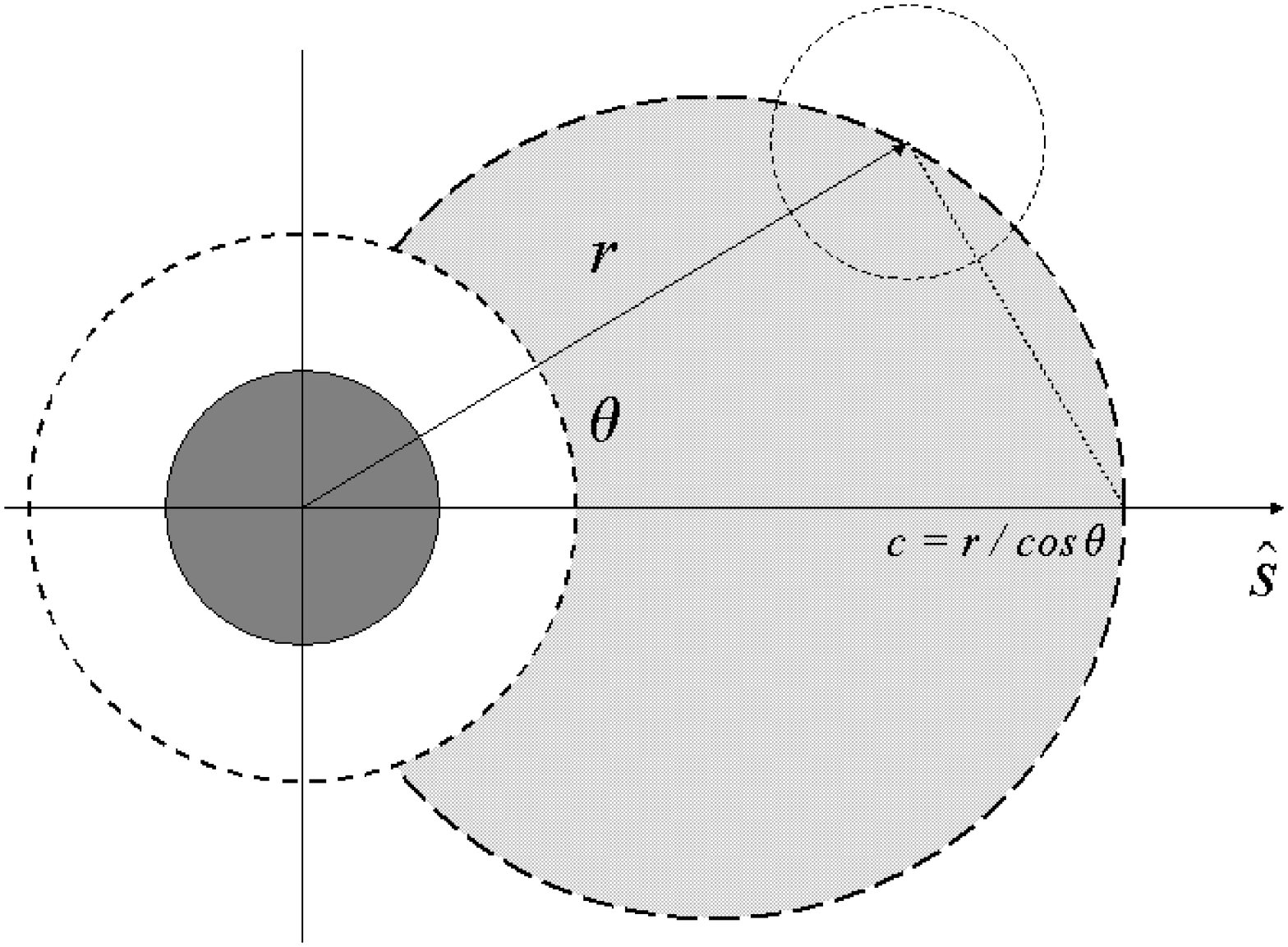}
}
\caption{ (a)
  The limit of the Voronoi cell of particle $i$ in the direction $\s$
  is $r_{ij}/2 \cos \theta_{ij}$. Then the Voronoi volume is
  proportional to the integration of $(r_{ij}/2 \cos \theta_{ij})^2$
  over $\s$ as in Eq.  (\ref{vor1}).
  (b) The particle contributing to the Voronoi volume along $\s$ is
  located at $(r,\theta)$.  The dark gray region is the considered
  grain ($r<R$), and in white the excluded zone for the center of any
  other grain ($r<2R$). For a given $c=r/\cos \theta$, the light grey
  area is the region of the plane $(r',\theta')$ where $r' / \cos
  \theta' < c $.}
\label{voronoi}
\end{figure}

The volume function is the key of the system's statistics: its flat
average over all jammed configurations determines the total volume.
The most natural way of dividing the system is called the Voronoi
diagram, which can be seen in Fig. \ref{voronoi}a.  Each grain's
region is the part of the space closer to this grain than to any
other, so that the volume is clearly additively partitioned. The major
drawback of this construction is that, so far, there was no analytical
formula for the Voronoi volume of each cell, such that attempts have
been made to use other constructions \cite{ball_blu}.

The Voronoi volume of a particle $i$ can be written as:

\begin{equation} \label{vor1}
{\cal W}_i^{\rm vor} = \frac{1}{2} \oint (\min_{\s\cdot\drij
> 0}\frac{\rij}{2\s\cdot\drij})^2 ds,
\end{equation}
where $\vrij$ is the vector from the position of particle $i$ to $j$,
the
average is over all the directions $\s$ forming an angle $\theta_{ij}$
with $\vrij$ as in Fig. \ref{voronoi}a. This formula has a simple
interpretation depicted in Fig. \ref{voronoi}a.  For consistency of
notation with previous work, we will use the words "volume" and
"surface" in 2d, although they correspond to "surface" and "length"
respectively.



The volume function defined in terms of the particle coordinates is of
no use, since it does not permit the calculation of the partition
function.  To solve this problem, we calculate an average free volume
function based on the environment of the particle, referring to a
coarse-graining over a certain mesoscopic length scale.
We assume a probability distribution for the positions of the nearest
neighbors as well as the other particles. After averaging over the
probability distribution we obtain an average mesoscopic free volume
function representing quasiparticles in the partition
function. Considering isotropic amorphous packings allows for removal
of the orientational averaging.


\section{Probability distribution of volumes}

Using the notation of Fig. \ref{voronoi}b,
we see that the microscopic volume function is entirely defined by the
parameter $c = \min[r / \cos \theta]$.
The calculation of the average free volume function, $w$, requires
knowledge of the probability distribution of this parameter. That
is:
\begin{equation}
  \label{3-free_vol}
  w \equiv \langle {\cal W}^{\rm vor}_i\rangle / V_g- 1 =
  - \int_{c=1} ^
  \infty (c^2-1) \frac{dP_>}{dc} \ dc,
\end{equation}
where $P_>(c)$ represents the inverse cumulative distribution,
i.e. the probability that all balls verify $r_{ij}/\cos \theta_{ij}
>c$, which is calculated under the following hypothesis:

(1)   The   cumulative  distribution   $P_>(c)$   is   made  of   two
contributions, one  of the  "contact" balls (\emph{i.e.}  touching the
considered grain),  called $P_>^C$, and one of  the other (background)
balls, $P_>^B$.
These probabilities can be understood as the probabilities of a
particle in contact (resp. background) for being situated outside the
grey zone on Fig. \ref{voronoi}b, and therefore not contributing to
the Voronoi volume $c^2$.




(2) The probability distributions are those of a large number of particles
at a given density, and of negligible size giving rise to
Boltzmann-like distributions: $$P_>^B(c)=\exp(-\rho V^*(c))$$ and
$$P_>^C(c)=\exp(-\rho_S S^*(c).$$
Here $V^*$ and $S^*$ represent a free volume and surface respectively
towards $c$, \emph{i.e}. $V^*(c)=\int \Theta(c-\frac{r}{\hat{r}\cdot
  \hat{s}}) d\vec{r}$ and $S^*(c)=\oint \Theta(c-\frac{1}{\hat{r}\cdot
  \hat{s}}) ds$ where the integrals cover respectively the space and
the unity sphere. The densities, $\rho(w)$ and $\rho_S(z)$, are mean
free-volume and free-surface densities, respectively.

In 2d, the volume of the grain (with $2R=1$) is $V_g = \pi /4$. The
free volume density (inverse of the free volume per particle) is $\rho(w)
= N/(N V_g \phi^{-1} - N V_g) = 1/(V_g w)$. Then,
$$ V^*(c) = \Big [\frac{c^2}{2} -1 \Big ] \arccos(1/c) + \frac{c}{2}
\sqrt{1-\frac{1}{c^2}}.$$ The main assumption here is that the packing
structure is uniform, thus the pair distribution function is assumed
to be a delta function at contact plus a constant for larger
distances.  This assumption is an oversimplification, and more
realistic background could be considered, such as peaks in the pair
distribution at the next nearest neighbor sites.

For the surface contribution, we have: $$S^*(c) = 2
\int_{0}^{\arccos(1/c)} d \theta = 2 \arccos(1/c).$$

\begin{figure}[h]
\centerline{(a)
\includegraphics[width=0.55\linewidth]{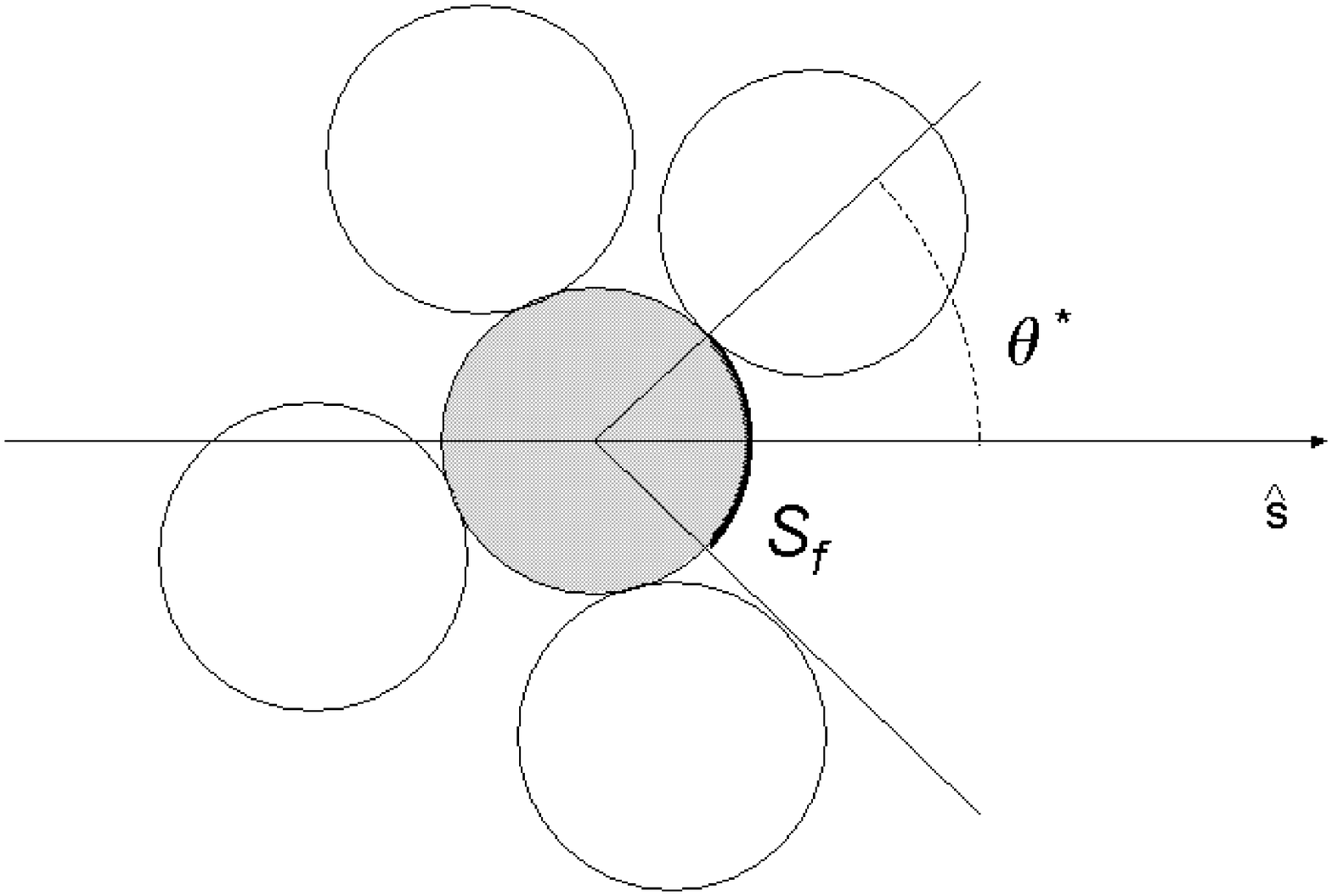}
}
\centerline{(b)
\includegraphics[width=0.55\linewidth]{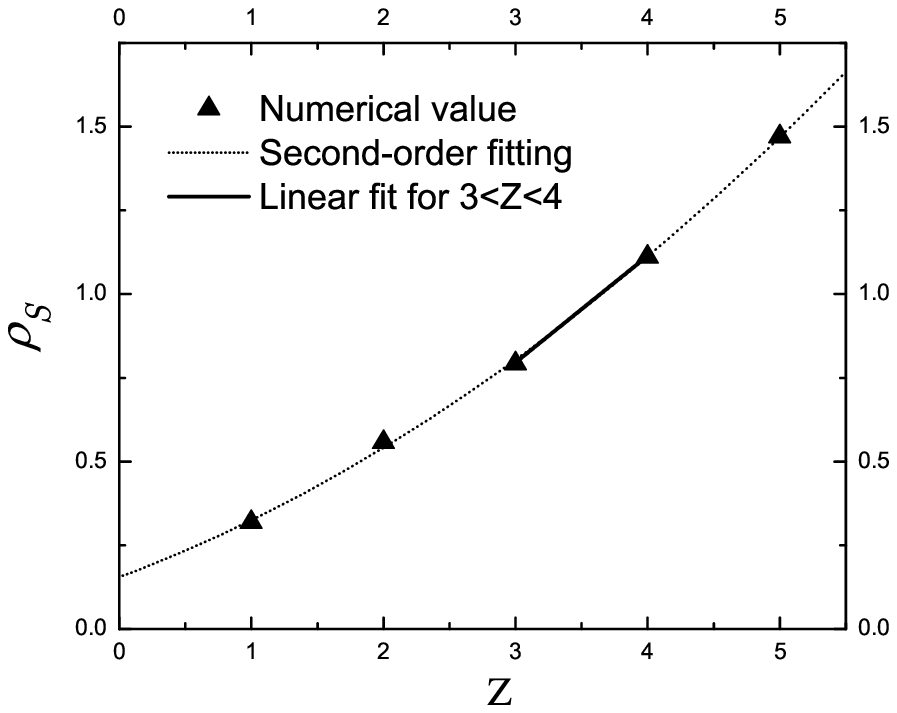}
}
\caption{(a) Free surface, in an example of $z=4$
  contact balls, defined by the angle of the closest grain to
  $\hat{s}$, with $\langle S\rangle=2 \theta^*$. (b) Simulation
  results for $1 \leq z \leq 5$, with a second-order polynomial
  fitting, and linear fitting for $3 \leq z \leq 4$ :
  $\rho_S=(z-0.5)/\pi$}
\label{free-surface}
\end{figure}

The surface density $\rho_S=1/\langle S \rangle$ is the inverse of the
average surface left free by $z$ contact balls (see Fig.
\ref{free-surface}a). As a rough approximation, one can assume it is
proportional to $z$, but because of the size of one ball, there is an
"excluded-surface" effect, so that the exact value is determined by
numerical simulations. It consists in setting sequentially and
randomly $z$ non-overlapping circles of radius 1 at the surface of the
unity circle (Fig. \ref{free-surface}a). The closest ball to the
considered direction $\hat{s}$ defines the free angle.  The free
surface is then twice this angle. Its average value is the mean
free-surface $\langle S \rangle$.

Results are shown in Fig. \ref{free-surface}b. Important deviations
from the linearity in $z$ are not surprising, since each contact ball
occupies an important surface ($z_{max}=6$ in 2d), and strong
finite-surface effects are expected. For $z=5$, in around $41\%$ of
the trials, the fifth ball cannot be set because there is not enough
space, and we take into account only the $59\%$ remaining trials.

For $3 \leq z \leq 4$, we will use the linear dependency $\rho_S(z) =
\frac{z-0.5}{\pi}$ as fitted in Fig. \ref{free-surface}.  Obviously a
fitting for $1 \leq z \leq 5$ would be of higher order, but in our
range, the error is insignificant compared to other approximations of
the model.

(3) The cumulative distributions are not independent. The assumption
    that the surface and volume terms do not overlap seems to be an
    abusive approximation in 2d. This is not the case in 3d as shown
    in \cite{swm}. Indeed, for higher dimensions the probabilities are
    expected to become independent, but in the case of 2d a new
    solution has to be worked out which considers the correlations
    between the contact and background term.

\begin{figure}[h]
\centerline{(a) \includegraphics[width=0.55\linewidth]{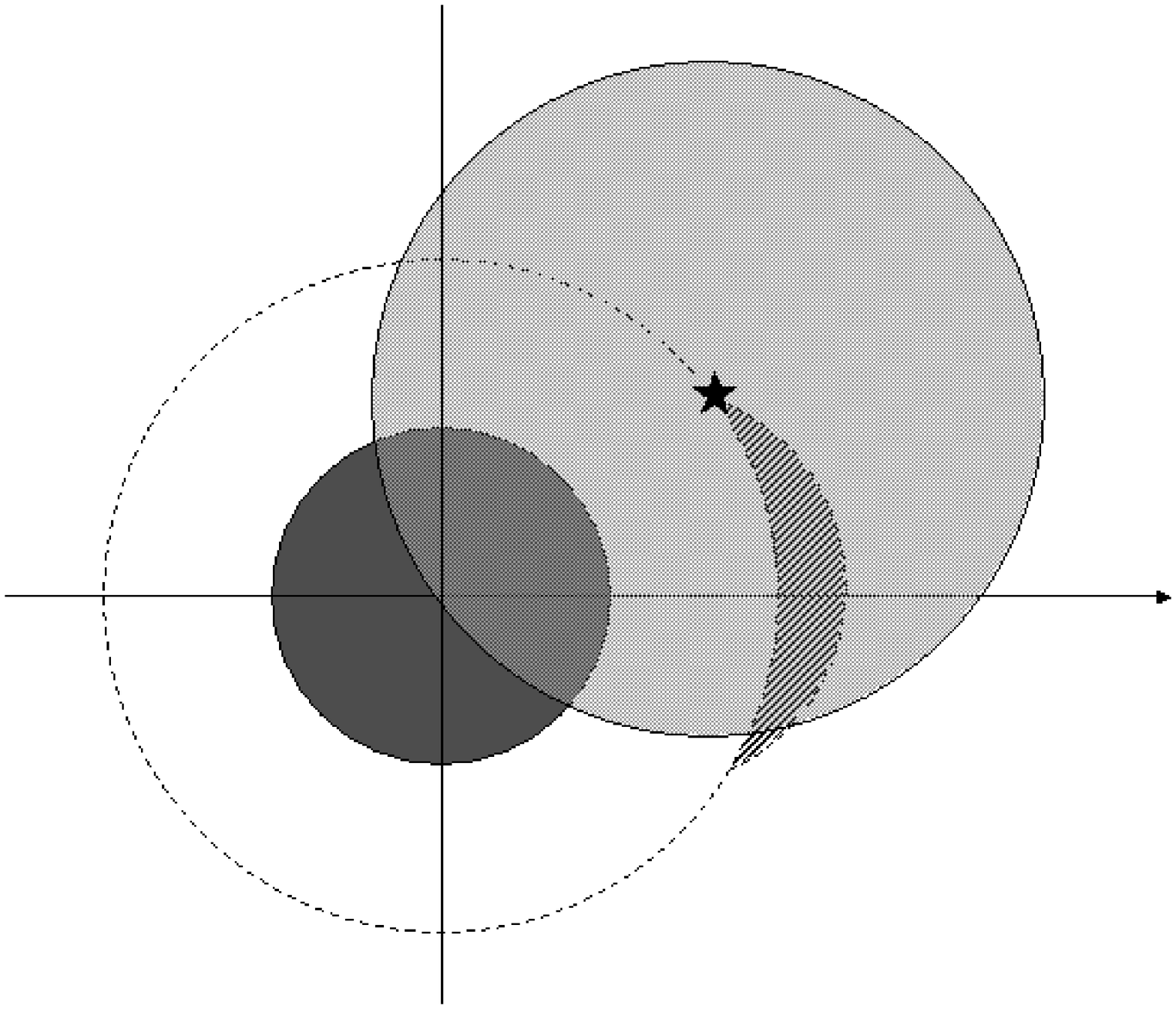}}
\centerline{(b)
\includegraphics[width=0.55\linewidth]{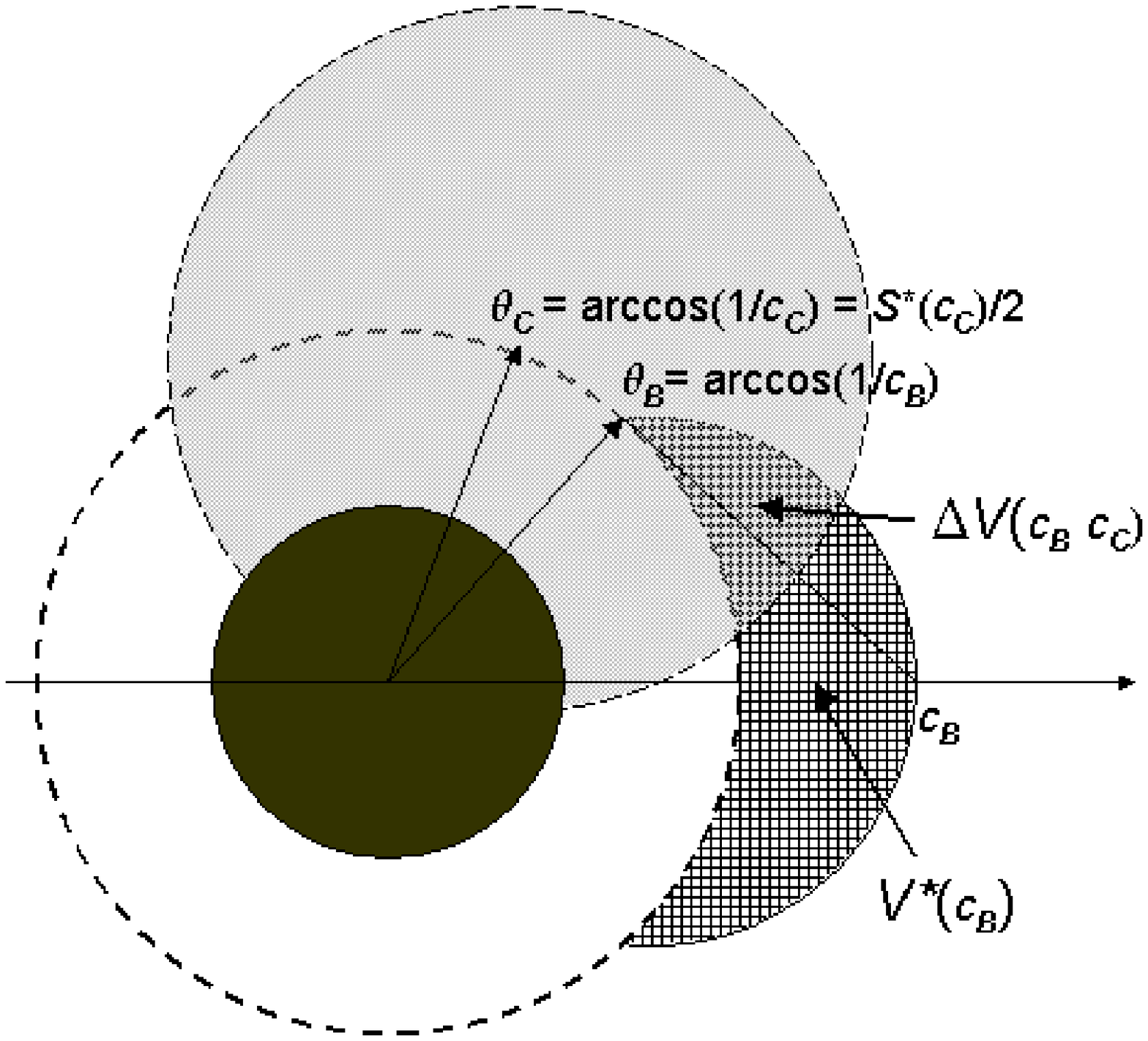}
}\caption{{\small (a) For $c=1.2$, the star is the center of the
    closest contact-grain, occupying the circular region printed with
    a pattern. The free volume is printed in grey. It is almost
    completely overlapped by the surface contribution. (b) The closest
    contact-ball depends on $c_C$, the free volume on $c_B$, and
    $\Delta V$ on both. Here it is the intersection between the region
    in light grey and the patterned region. }}
\label{overlap}
\end{figure}

    To illustrate this point, Fig. \ref{overlap}a shows the considered
    grain, the free volume $V^*(c)$ and the circle occupied by the
    closest "contact-grain" (\emph{i.e.} the excluded zone for the
    center of any other grain because of its presence) for a value of
    $c=1.2$. In fact, the values of $c<1.2$ are contributing for
    $94\%$ of the distribution if we neglect the overlap of contact
    and background grains.  As we see, the free volume is mostly
    covered up by this contact-grain, and the non-overlapping
    hypothesis used in \cite{swm} appears obviously wrong. This
    statement is confirmed by the calculation of the RCP density:
    with the non-overlapping hypothesis, the calculation provides a
    value of $\phi_{RCP}=0.89$, to be compared with the reported value
    of $0.82$. The nearer grains are exceedingly taken into account.

Therefore, the volume term is modified by substituting the free volume
$V^*$ by $V^*-\Delta V^*$ which represents the free volume minus the
part occupied by the closest surface grain.  The meaning of this
change is that the contributions are no longer independent, and depend
on two parameters $c_B$ and $c_C$.  The distribution $P_>(c)$ is the
probability that both $c_C$ and $c_B$ be higher than $c$.  Figure
\ref{overlap}b shows the overlap of the contact grain parameterized by
$c_C$ and the background volume parameterized by $c_B$, defining
$\Delta V^* (c_B, c_C)$. The analytical formula of $\Delta V^*$ is
determined by geometrical calculations.


The probability density is
$$P(c) = -dP_>/dc = -P_>^C(c) \cdot dP_>^B/dc - P_>^B(c) \cdot
dP_>^C/dc.$$
 The meaning of the latter equality is that, to realize $c$,
we must have either $c_C$ or $c_B$ equal to c, and the other
higher. The background probability depends on $c_C$: $$P_>^B(c_B|c_C)
= \exp[- \rho (V^*(c_B)-\Delta V^* (c_B,c_C))],$$ and $$P^B(c_B|c_C) =
-\frac{d }{dc_B}P_>^B(c_B|c_C).$$ Then, $$P(c) = \int_{c_C=c}^{\infty}
P^C(c_C) P^B(c|c_C) dc_C + P^C(c)P_>^B(c|c)= $$ $$ =
- \frac{d}{dc}
\int_{c_C=c}^{\infty} P^C(c_C) P_>^B(c|c_C) dc_C =-
\frac{d}{dc}P_>(c).$$


From (\ref{3-free_vol}), we integrate by parts using the latter equality. The
boundary term $[(c^2-1)P_>(c)]$ vanishes, since the limits of
integration correspond to $c=1$ and $c \rightarrow \infty$, with $P(c
\rightarrow \infty) = 0 $. We obtain for the average
volume function from Eq. (\ref{3-free_vol}):
\begin{equation}
  w = 2\int_{c=1}^\infty c \int_{c}^\infty  \frac{dP_>^{C}}{dc_C}
  \exp [ -\rho(w) \big ( V^*(c)-\Delta V^* (c,c_C) \big ) ] dc_C dc
\label{self}
\end{equation}
with $P_>^{C}(c_C)=\exp[-\rho_S(z) S^*(c_C)]$.



\section{Free volume function}

\begin{figure}[h]
\centerline{(a) \includegraphics[width=0.55\linewidth]{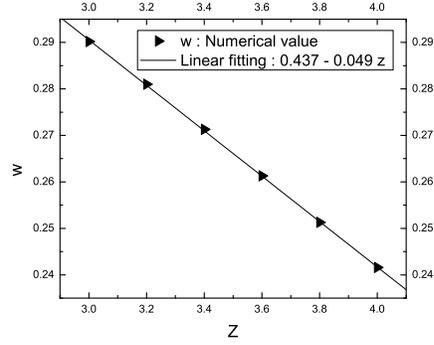}}
\centerline{(b) \includegraphics[width=0.55\linewidth]{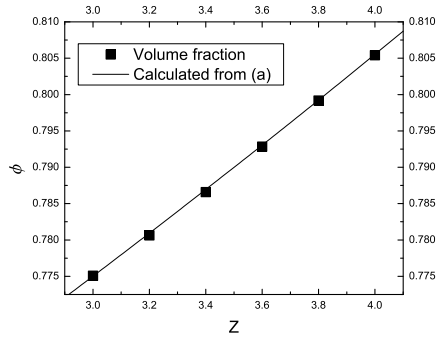}}
\centerline{(c) \includegraphics[width=0.55\linewidth]{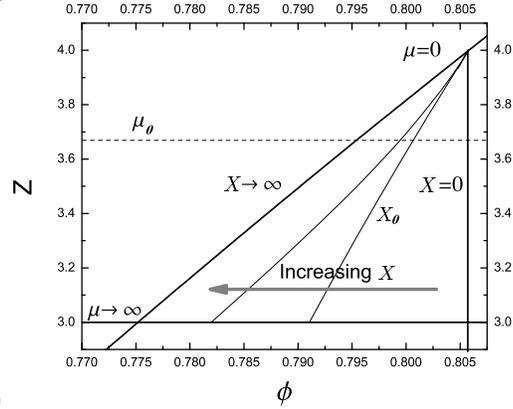}
}
\caption{ (a) $w(z)$ curve, with a linear fitting: $w(z)=0.437-0.049z$.
(b) Volume fraction according to the second Eq. (\ref{4-w_Z}).
  (c) Prediction of the model.  The thick curves are the limit of the
  diagram at $X=0$ and  $X \rightarrow \infty$.
We show several curves of constant
  compactivity, $X$.
The curves are plotted for (from right to left) :
    $X=5.10^{-3}$, $X=10^{-2}$ and $X=10^{-1}$.
  The horizontal lines are both constant $Z$ given by an arbitrary
  $\mu_0$ (dotted) and $\mu \rightarrow \infty$ (thick inferior limit of
  the diagram).}
\label{w}
\end{figure}

Equation (\ref{self}) is a self-consistent equation to obtain $w(z)$,
which cannot be solved exactly, therefore a numerical integration of
(\ref{self}) is necessary to obtain $w$ vs $z$. For various values of
$z$, we integrate Eq. (\ref{self}) numerically, and we then calculate
a fitting of the results (Fig. \ref{w}a).  We obtain the free volume
function and the local density $\phi_i^{-1}=w+1$ (Fig. \ref{w}b):
\begin{equation}\label{4-w_Z}
w(z)=0.437-0.049z, \ \ \ \ \ \ \phi_i(z)= \frac{1}{1.437-0.049z}.
\end{equation}





Generally speaking, we would expect $w$ to be roughly proportional to
$1/z$, with $w \rightarrow 0$ when $z \rightarrow \infty$. However,
the statement $z \rightarrow \infty$ has little meaning when we plot a
figure for $3<z<4$ and the "infinite" (maximal) value of $z$ is 6.

\section{Statistical mechanics}


Equation (\ref{4-w_Z}) plays the role of a "Hamiltonian" of the system.
Each jammed configuration corresponds to some "volume level" in
analogy with energy levels in Hamiltonian systems.
From the formal analogy with classical formulas, the canonical
partition function is \cite{sirsam}:

\begin{equation}\label{4-part1}
\mathcal{Z} (Z,X) = \int g(w) e^{-w/X} dw,
\end{equation}
where
$X$ is the (reduced) compactivity (normalized by the volume of the
spheres) and $g(w)$ is density of states for a given volume $w$. We
remind that the compactivity is the equivalent of temperature in the
Edwards statistics, and it is a measure of the system's looseness.


Since the volume $w$ is now directly related to coordination number
$z$ through Eq. (\ref{4-w_Z}), we can compute $g(w)$ by replacing
variable,
$g(w) = \int P(w|z) g(z) dz,$
where $P(w|z)$ is the conditional probability and $g(z)$ is the
density of states for given $z$ \cite{swm}.

At this point a distinction has to be emphasized:
we refer to $z$ as the geometrical coordination number since it is
purely defined by the particle positions.  On the other hand,
there is the mechanical coordination number, $Z$, defined by those
geometrical contacts that carry a non-zero force. $Z$ is then
defined by the mechanical constraint leading to the isostatic
condition. A packing is isostatic when the number of contact
forces equals the number of force and torque balance equations
\cite{swm}. For example, for a packing of $N$ infinitely rough
particles in $d$ dimensions, each mechanical contact carries one
normal force and $d-1$ tangential force components, and for each
particle there are $d$ force balance equations and
$\frac{1}{2}d(d-1)$ torque balance equations. The isostatic
condition requires that $\frac{1}{2}dNZ = dN +
\frac{1}{2}d(d-1)N$, or $Z=d+1$. On the other hand, for
frictionless packings, frictional forces or tangential forces do
not exist and the torque balance equations are not taken into
account. The isostatic condition in this case leads to the
relation $Z = 2d$. For a system with a finite interparticle
friction $\mu$, $Z(\mu)$ interpolates between both limits
\cite{kertesz,swm}. It should be noted that the above calculations
of the isostatic condition are based on the mechanical
coordination number $Z$, rather than the geometrical coordination
number $z$, because a geometrical contact does not necessarily
provide a mechanical constraint. For instance, two particles are
free to rotate with respect to each other if the contact between
them is geometrical and does not carry any tangential forces.



Obviously, $z$ must be larger than $Z$ for the mechanical condition to
be satisfied. The different volume levels of a packing can be
understood in the following way: the friction coefficient sets a
mechanical constraint on $Z$, but the system can explore all
geometrical levels $z>Z$. Additionally, $z$ is bounded by the maximal
coordination number for a random packing, which is $2d=4$, since there
are $z/2$ constraints on the $d$ particle coordinates.
In relation with the discussion of isostaticity, it is believed that
above this value, the system is partially crystallized. Therefore we
obtain: $g(w) = \int_Z^4 P(w|z) g(z) dz.$
Since $w(z)$ from Eq. (\ref{4-w_Z}) is a coarse-grained free volume
and independent of the microscopic positions of particles, we have
$P(w|z) = \delta ( w - w(z) )$.



The expression of the density of states is obtained by considering that
the states are collectively jammed.  Therefore, the space of
configuration is discrete since we cannot continuously obtain one
configuration from another. Assuming a typical distance between
configurations as $h_z$, we obtain $g(z) \propto (h_z)^{z}$, the
exponent $z$ arising
since there are $z$ position constraints per particle in the jammed
state compared to the free (gas) state.
Such a formula is analogous to the factor $h^{-d}$ for the density of
states in traditional statistical mechanics, where $h$ is the Planck
constant, which arises because of the uncertainty principle,
\emph{i.e.} because of the discreteness of the elementary volume of
phase space.


Substituting into Eq.  (\ref{4-part1}), we get :
\begin{equation}\label{4-part_2}
  \mathcal{Z}(Z,X) = \int_Z^4 (h_z)^z e^{-w(z)/X} dz
\end{equation}






To establish the maximum and minimum densities, we consider the limits
of zero and infinite compactivity, respectively.
The ground state of jammed matter, is analogous with the limit
$T\rightarrow 0$. The only accessible state is $z=4$, corresponding to
the random close packing. For this state, from Eq. (\ref{4-w_Z}) we
get a fixed value of the volume fraction for any coordination number
$Z \in [3,4]$, $\mu \in [0,\infty]$:
\begin{equation}\label{4-RCP}
\phi_{\rm rcp}(Z) = \frac{1}{1.437-0.049\times4} \ \approx \ 0.806,
\end{equation}

The lower density appears for $X \to \infty$ when the Boltzmann factor
is unity in Eq. (\ref{4-part_2}), and we obtain the densities of RLP
(assuming $h_z\ll 1$):

$$  \phi_{\rm rlp}(Z) = \frac{1}{\mathcal{Z}(Z,\infty)} \int_Z^4
    \frac{1}{1.437-0.049z}(h_z)^z dz \approx$$
\begin{equation}
\approx    \frac{1}{1.437-0.049Z}, \ \ Z \in [3,4].
\label{4-RLP}
\end{equation}


This leads to the diagram in the plane $(\phi,Z)$ plotted in
Fig. \ref{w}c defining the possible jammed configurations.  On
the right part of the vertical line defined by Eq. (\ref{4-RCP}), no
disordered packing can exist. To the left of Eq. (\ref{4-RLP}), the
packings are not mechanically stable.

Between these two lines, we plot the lines of constant finite
compactivity. For a finite value of the compactivity, the equation
$\phi(Z)$ is calculated by numerical integration.
In the figure, three curves are plotted, respectively $X=5.10^{-3}$,
$X=10^{-2}$ and $X=10^{-1}$ for $h_z=0.01$. The compactivity increases
from the right ($X=0$) to the left ($X \to \infty$). The limit
$\mu\to\infty$, $Z\to 3$ defines the lowest RLP density value which is
predicted:
\begin{equation}
  \phi_{\rm rlp}^{\rm min} = \phi_{\rm rlp}(Z=3) =
  \frac{1}{1.437-0.049 \times 3} \approx 0.775
\end{equation}

The value of $\phi_{\rm rlp}$ depends on the mechanical coordination
number, contrarily to the value of $\phi_{\rm rcp}$.
The shape of the diagram is similar in 3d \cite{swm}, and this is in
agreement with the wide range of reported values for RLP, in contrast
with RCP \cite{berryman}. On a horizontal line given by a system with
fixed $Z$, packings of different volume fractions can be achieved by
applying different quench rates or compression speeds during the
preparation protocol. Slow compressions achieve loose packings (and
high compactivities). The obtained predictions for the density of RCP
are close to the experimental values while we predict the existence of
a RLP density.


\section{Summary}

In summary, we have used a model of volume fluctuations to develop a
statistical mechanics of granular matter in 2d. From a quantitative
point of view, we have seen how it lies on several approximations,
that can appear too rough.  The main difference with the 3d case is
the need of taking into account properly the correlations in the
probability distribution of volume through the consideration of point
(3) above.  Indeed, if we take $P_{>}^{\rm C}$ and $P_{>}^{\rm B}$ to
be independent as considered in \cite{swm} we find $\phi_{\rm
rlp}=0.84$ and $\phi_{\rm rcp}=0.89$, both values above the
experimental value of RCP.
The results, although not allowing exact predictions, are situated in
the right order of magnitude for the limiting volume fractions. Due to
the several approximations of the theory, the resulting limiting
densities have to be considered as bounds to the real
values. Improvements can be achieved by taking into account the size
and shape of the disks, as well as exact enumeration to calculate
$P_>(c)$, which can be done at least to a prescribed coordination
shell of particles, in a brute force analysis analogous to the Hales
proof of the Kepler conjecture, currently being
undertaken. Altogether, the present framework seems to be successful
in describing at least qualitatively the general features of jammed
granular matter in 2d providing evidence of the existence of RCP and
RLP and their density value.  These results suggest that a putative
ideal glass transition may also exist in frictionless hard disk as
discussed in \cite{parisi}.


\pagebreak[4]

\end{document}